# Computing Riemann zeros with light scattering

Sunkyu Yu[1†], Xianji Piao[2§], and Namkyoo Park[3*]

[1]Intelligent Wave Systems Laboratory, Department of Electrical and Computer Engineering, Seoul National University, Seoul 08826, Korea

[2]Wave Engineering Laboratory, School of Electrical and Computer Engineering, University of Seoul, Seoul 02504, Korea

[3]Photonic Systems Laboratory, Department of Electrical and Computer Engineering, Seoul National University, Seoul 08826, Korea

E-mail address for correspondence: [†]sunkyu.yu@snu.ac.kr, [§]piao@uos.ac.kr, [*]nkpark@snu.ac.kr

**Abstract**

Finding hidden order within disorder is a common interest in material science[1,2], wave physics[3,4], and mathematics[5]. The Riemann hypothesis, stating the locations of nontrivial zeros of the Riemann zeta function[6], tentatively characterizes statistical order in the seemingly random distribution of prime numbers. This famous conjecture has inspired various connections with different branches of physics[7], recently with non-Hermitian physics[8,9], quantum field theory[10], trapped-ion qubits[11], and hyperuniformity[12,13]. Here we develop the computing platform for the Riemann zeta function by employing classical scattering of light. We show that the Riemann hypothesis suggests the landscape of semi-infinite optical scatterers for the perfect reflectionless condition under the Born approximation. To examine the validity of the scattering-based computation, we investigate the asymptotic behaviours of suppressed reflections with the increasing number of scatterers and the emergence of multiple scattering. The result provides another bridge between classical physics and the Riemann zeros, exhibiting the design of wave devices inspired by number theory.

**Introduction**

As an abstract language of physics, mathematics allows for linking seemingly unrelated phenomena in different domains of physics. For example, the mathematical similarity in governing equations enables analogies between classical and quantum phenomena[14-16]. Statistical measures classify distinctly disordered materials into the same material phase[1-4]. The concept of topology provides a common design strategy for implementing robust systems across different domains[17-19]. Similarly, the Riemann hypothesis[6] is a representative example that bridges classical, quantum, and relativistic physics through number theory[7].

For the Riemann zeta function $\zeta(s) = \sum_{n=1}^{\infty} \frac{1}{n^s}$ with a complex variable $s = \sigma + it$, where $\sigma$ and $t$ are real, this cornerstone conjecture claims that the locations of the nontrivial zeros for $t \neq 0$ are restricted to $\sigma = 1/2$. The most important result of understanding the locations of the nontrivial Riemann zeros would be the quantification of prime number distribution due to its relationship with the prime counting function, which forms the basis for number theory[5,20]. Efforts to connect the conjecture to physics problems have been focused on finding $\zeta(s)$ in various systems: chaotic resonators[7], parity-time-symmetric Hamiltonians[8,9], relativistic scattering[10], and trapped-ion qubits[11]. The most famous approach is the Hilbert-Pólya conjecture[21], which suggests that the nontrivial Riemann zeros correspond to real eigenvalues of a Hermitian operator.

Material science is another branch exploring the implications of the Riemann hypothesis. Due to the similarity between material classification and number theory in discovering hidden order within highly randomized patterns, the intermediate regime between order and uncorrelated disorder has attracted substantial attention in exploring the hypothesis. For example, there has been an attempt to understand the locations of the Riemann zeros using quasicrystal patterns[22,23]. The distribution of prime numbers has also been examined through the concept of hyperuniformity[24-26], which reveals the suppression of long-range fluctuations in the pattern of prime numbers[12,13].

Here, we revisit the classical theory of light scattering to develop wave-based computing platforms for the Riemann zeta function through engineered scatterers. We propose the concept of the $\sigma$-Riemann scatterer, which is composed of the logarithmically spaced one-dimensional (1D) semi-infinite scatterers with $\sigma$-power-law-decayed scattering amplitudes. We demonstrate that the Riemann hypothesis corresponds to the complete reflectionless condition only at the (1/2)-Riemann scatterer under the first-order Born approximation, where the locations of the Riemann zeros denote the zero-reflection wavelength of light. To exploit this reflectionless condition for optical devices, we estimate the valid range of highly suppressed reflection at the Riemann-zero wavelengths, by examining a finite number of scatterers and multiple scattering beyond the Born approximation. The result offers insights into the non-Hermitian generalization of the Hilbert-Pólya conjecture and the design of reflectionless optical structures using the hyperuniformity with reciprocal-space shifts.

## Results

**Riemann zeta functions in light scattering**

To investigate the relationship between light scattering and the Riemann zeta function, we review 1D multiple scattering theory[27]. Consider a wave equation $d^2\psi/dx^2 + [V_0 + V(x)]\psi = 0$, where $\psi$ is a wavefunction, and $V_0$ and $V(x)$ are a constant and a spatially-varying potential, respectively. Although the potentials in our example correspond to the landscape of permittivity in nonmagnetic optical media[15,28], the equation is applicable to describe quantum states under the single-particle approximation[29], and classical oscillations of matter, such as acoustic[30] and elastic[31] waves. The profile of $V(x)$ determines material inhomogeneity.

Motivated by the semi-infinite summation of $\zeta(s) = \sum_{n=1}^{\infty} \frac{1}{n^s}$, we consider a material that is asymptotically homogeneous as $x$ approaches negative infinity: $V(x \to -\infty) = 0$. Scattering from the material is examined with the planewave incidence from $x \to -\infty$, $\psi_I = \psi_0 \exp(ik_0 x)$, which is the solution of the Helmholtz equation $d^2\psi/dx^2 + V_0\psi = 0$, where $V_0 = k_0^2$ and $k_0 = 2\pi/\lambda \geq 0$ for the wavelength $\lambda$. With the scattering wavefunction $\psi_S$, the total wavefunction $\psi$ is determined by the 1D Lippmann-Schwinger equation[27], as $\psi = \psi_I + \psi_S = \psi_I + \frac{1}{4\pi}\int_{-\infty}^{+\infty} V(x')\psi(x')G(x,x')dx'$, where $G(x,x') = (2\pi i/k_0)\exp(ik_0|x-x'|)$ is the 1D Green's function satisfying $d^2G(x,x')/dx^2 + V_0 G(x,x') = -4\pi\delta(x-x')$ for the Dirac delta function $\delta(x)$. By defining the scattering operator $M$ as

$$(M\varphi)(x) = \frac{i}{2k_0}\int_{-\infty}^{+\infty} V(x')\varphi(x')e^{ik_0|x-x'|}dx', \tag{1}$$

for an arbitrary wavefunction $\varphi(x)$, $\psi$ can be expressed as the Born series[27], $\psi = \psi_I + \sum_{m=1}^{\infty} \psi_{S,m}$, where $\psi_{S,m} \triangleq M^m \psi_I$ is the $m$th-order scattering wavefunction.

From the above review of the 1D scattering theory, we derive the Riemann zeta function $\zeta(s)$ in the first-order scattering $\psi_{S,1}$, by tailoring the spatial inhomogeneity of the potential $V(x)$. The inhomogeneous potential is implemented with an array of $N$ point scatterers, as $V(x) = V_N(x) \triangleq \sum_{n=1}^{N} d_n \delta(x - x_n)$, where $d_n$ and $x_n$ are the scattering amplitude and position of the $n$th scatterer, respectively, $N$ is the number of scatterers, and $x_n \leq x_{n+1}$ for all $n$. Among the multifaceted forms of $\zeta(s)$, we focus on implementing the following form obtained from the Dirichlet eta function[32,33]:

$$\zeta(s) = \frac{1}{1-2^{1-s}} \sum_{n=1}^{\infty} \frac{(-1)^{n+1}}{n^s}, \tag{2}$$

to guarantee numerical convergence in the critical strip $\{s\in\mathbb{C}: 0 < \text{Re}[s] = \sigma < 1\}$, which was already proved to possess all the nontrivial zeros[20]. To derive Eq. (2) in the 1D scattering formulation, we define the $\sigma$-Riemann scatterer (Fig. 1a)—an array of $N$ scatterers having the logarithmically spaced distribution $x_n = x_0 \log(n)$ and the alternating $\sigma$-power-law scattering amplitudes $d_n = (-1)^{n+1} d_0 n^{-\sigma}$, as

$$V(x) = d_0 \sum_{n=1}^{N} \frac{(-1)^{n+1}}{n^\sigma} \delta(x - x_0 \log(n)), \qquad (3)$$

where $x_0$ and $d_0$ are real- and complex-valued coefficients, respectively. The resulting first-order scattering in the region between the $p$th and $(p+1)$th scatterers is obtained with Eq. (1), as follows (Supplementary Note S1):

$$\psi_{S,1}(x; x_p \leq x < x_{p+1}) = \frac{id_0 \psi_0}{2k_0} \left[ \left( \sum_{n=1}^{p} \frac{(-1)^{n+1}}{n^\sigma} \right) e^{+ik_0 x} + \left( \sum_{n=p+1}^{N} \frac{(-1)^{n+1}}{n^{\sigma - 2ik_0 x_0}} \right) e^{-ik_0 x} \right]. \qquad (4)$$

The comparison between the first-order scattering $\psi_{S,1}$ and multiple scattering $\psi_S$ is shown in Figs. 1b and 1c for small and large values of $d_0$, respectively, which exhibit the validity of Eq. (4) for small $d_0$ (Supplementary Note S2 for the calculation of multiple scattering).

Equation (4) provides insight into the connection between the first-order scattering and the Riemann zeta function, leading to the following transmitted ($p \to \infty$) and reflected ($p = 0$) waves from the semi-infinite ($N \to \infty$) $\sigma$-Riemann scatterer:

$$\begin{aligned} \psi_T(x; x \to \infty) &= \lim_{p \to \infty} \psi_{S,1}(x; x \geq x_p) = \frac{id_0 \psi_0 e^{+ik_0 x}}{2k_0} \left(1 - 2^{1-\sigma}\right) \zeta(\sigma), \\ \psi_R(x; x < x_1) &= \psi_{S,1}(x; x < x_1) = \frac{id_0 \psi_0 e^{-ik_0 x}}{2k_0} \left(1 - 2^{1-\sigma+2ik_0 x_0}\right) \zeta(\sigma - 2ik_0 x_0). \end{aligned} \qquad (5)$$

Equation (5) demonstrates that both the transmitted and reflected waves from the semi-infinite $\sigma$-Riemann scatterer are proportional to the Riemann zeta function $\zeta(s)$ under the first-order Born approximation. When considering a practical measurement setup and the exploration of nontrivial Riemann zeros with $s \in \mathbb{C}$, the measurement of the reflected wave is suitable for light-based computation of the Riemann zeta function. Notably, $\text{Im}[s] = -2k_0 x_0$ can be controlled by the optical wavelength $\lambda = 2\pi/k_0$. Figures 1d and 1e describe the $\lambda$-dependent amplitude and phase of the Riemann zeta function $\zeta(\sigma - 4\pi i x_0/\lambda)$, respectively, within the critical strip $0 < \sigma < 1$, exhibiting the Riemann-zero wavelengths ('×' symbols).

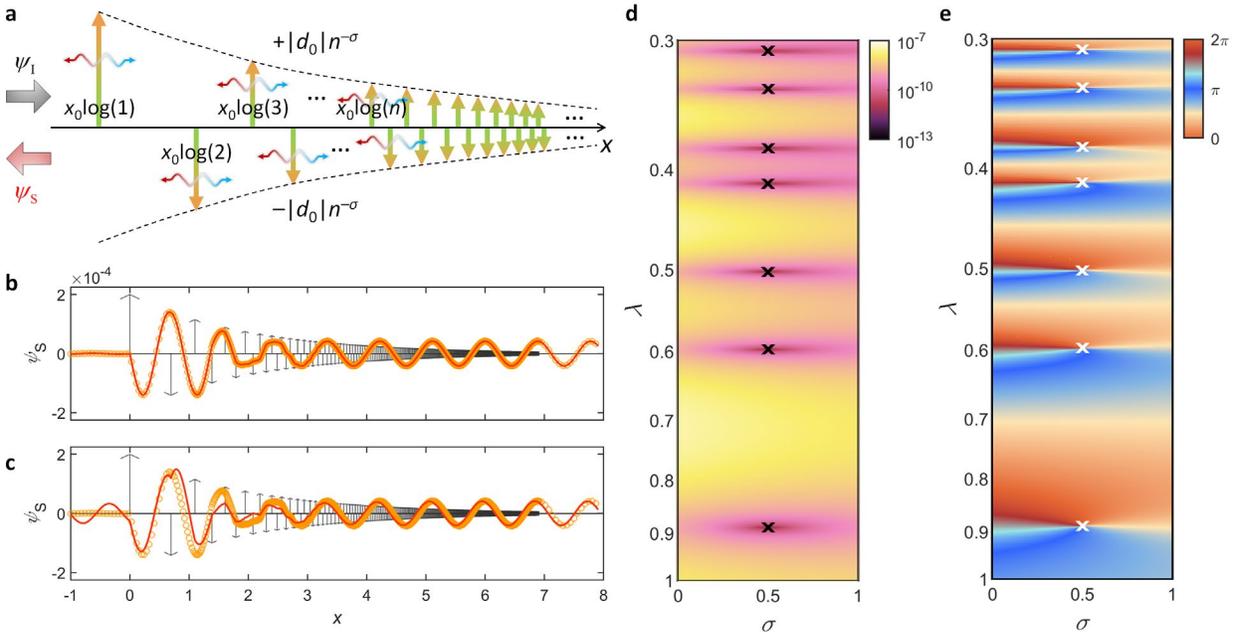

**Fig. 1. $\sigma$-Riemann scatterer. a,** A schematic of the $\sigma$-Riemann scatterer: a semi-infinite array of logarithmically spaced scatterers with a $\sigma$-power-law distribution of sign-alternating scattering amplitudes. The Riemann zeta function in the critical strip can be computed with the scattering wave $\psi_S$ for the incidence $\psi_I$. Red and blue curved arrows at each scatterer denote the scattering waves defined by the Green's function. **b,c,** Examples of scattering from the (1/2)-Riemann scatterer for $d_0 = 0.001$ (**b**) and $d_0 = 10$ (**c**). The arrows illustrate point scatterers. Red solid lines denote the multiple scattering $\psi_S$ calculated by the scattering matrix method[34] (Supplementary Note S2). Orange circles denote the first-order scattering $\psi_{S,1}$ obtained from Eq. (4). $N = 1000$, $x_0 = 1$ and $\psi_0 = 1$. **d,e,** The Riemann zeta function $\zeta(\sigma - 4\pi i x_0/\lambda)$ in the critical strip: amplitude (**d**) and phase (**e**) of $\zeta(\sigma - 4\pi i x_0/\lambda)$. The symbols '×' denote the first seven nontrivial Riemann zeros.

**Optically computed Riemann zeros**

Using the design of the Riemann scatterer, we revisit the Riemann hypothesis, which states that the nontrivial zeros of $\zeta(s)$ exist only at $\text{Re}[s] = \sigma = 1/2$. Because $\zeta(s^*) = [\zeta(s)]^*$, it is well known[20] that the nontrivial zeros $s^0 = 1/2 + it^0$ lead to the other zeros $s^0 = 1/2 - it^0$, which can be directly compared with $s = \sigma - 4\pi i x_0/\lambda$ in $\psi_R(x) \sim \zeta(s)$. Therefore, the Riemann hypothesis claims that the complete suppression of the first-order scattering from the $\sigma$-Riemann scatterer can be achieved only with the (1/2)-Riemann scatterer, which requires the scattering amplitudes $d_n = (-1)^{n+1} d_0 n^{-1/2}$ ($\sigma$ row in Table 1). The location of the nontrivial Riemann zeros $t^0$ determines the perfect reflectionless wavelength $\lambda^0 = 4\pi x_0/t^0$ ($t$ row in Table 1). The conjugate relation $\zeta(s^*) = [\zeta(s)]^*$ corresponds to electromagnetic reciprocity in light scattering, achieving the reflectionless condition for the incidence with the opposite-directional wavevector $-2\pi/\lambda^0$ ($s^*$ row in Table 1). We emphasize that higher-order scatterings $\psi_{S,m}(x) = M^{m-1}\psi_{S,1}(x) = 0$ ($m = 2, 3, \ldots, \infty$) are not completely suppressed because the complete suppression of $\psi_{S,1}(x)$ occurs only in the range of $x < 0$, and thus, nonzero $\psi_{S,1}(x \geq 0)$ with the integral operator $M$ leads to higher-order scattering with Eq. (1).

**Table. 1. Correspondence between the Riemann hypothesis and the first-order scattering.** The rows of the table indicate the comparison in terms of the real and imaginary parts of the variable $s$ and its conjugate $s^*$.

|  | Riemann hypothesis | First-order scattering from the $\sigma$-Riemann scatterer |
|---|---|---|
| $\sigma = \text{Re}[s]$ | $\sigma = 1/2$ for $\zeta(s) = 0$ | Scattering amplitudes $|d_n| = d_0/n^{1/2}$ for $\psi_R(x) = 0$ |
| $t = \text{Im}[s]$ | $t = t^0$ for $\zeta(s) = 0$ | Incident wavelength $\lambda^0 = 4\pi x_0/t^0$ for $\psi_R(x) = 0$ |
| $s^*$ | $\zeta(s^{0*}) = 0$ when $\zeta(s^0) = 0$ | Wavevector reciprocity ($\pm 2\pi/\lambda^0$) for $\psi_R(x) = 0$ |

Interpreting nontrivial Riemann zeros with the (1/2)-Riemann scatterer presents a unique approach to the Hilbert-Pólya conjecture[21] and its generalization into non-Hermitian physics[8,9]. Consider the decomposition of the Hermitian Hamiltonian $H$ for the (1/2)-Riemann scatterer into two subsystem Hamiltonians, which separate the forward ($H_F$) and backward ($H_B$) paths in the region where $x < 0$. Due to scattering, $H_F$ and $H_B$ are generally coupled, and therefore, each Hamiltonian itself is non-Hermitian with complex eigenvalues. However, under the first-order Born approximation, $H_F$ (and $H_B$) can exhibit real eigenvalues without the coupling to $H_B$ at the reflectionless wavelengths $\lambda^0$. Therefore, $H_F$ could serve as the non-Hermitian Hamiltonian for the generalized Hilbert-Pólya conjecture. The Riemann hypothesis states that such decomposition with real eigenvalues is possible only with the potential landscape of the (1/2)-Riemann scatterer.

The condition of nontrivial Riemann zeros with the (1/2)-Riemann scatterer also exhibits an intriguing connection to the concept of hyperuniformity[3,24-26,35], allowing for interpreting the (1/2)-Riemann scatterer as correlated disorder. For a 1D inhomogeneous material composed of point particles, $V(x) = V_N(x) \triangleq \sum_{n=1}^{N} d_n \delta(x - x_n)$, the structure factor $S(k)$ is given by $S(k) = \left|\sum_{n=1}^{N} d_n \exp(-ikx_n)\right|^2$. Hyperuniformity is defined by the asymptotic behaviour of the structure factor, $S(k \to 0) \to 0$ in the thermodynamic limit, which corresponds to the suppression of infinite-wavelength density fluctuations[3,24,26]. For the Riemann scatterer, the above definition of the structure factor leads to $S(k) = d_0^2 \left|\sum_{n=1}^{N} (-1)^{n+1} n^{-(\sigma + ikx_0)}\right|^2$, which approaches $S(k) = d_0^2 |\zeta(\sigma + ikx_0)|^2$ at $N \to \infty$ in the critical strip. Therefore, each nontrivial Riemann zero in the Riemann scatterer corresponds to the complete halt of the density fluctuation exactly at the wavelength determined by $\lambda^0 = 4\pi x_0 / t^0$, which can be considered the reciprocal-space translation of the perfect hyperuniformity from the infinity to $2k_0$.

**Evolving Riemann scatterer**

Although the correspondence between the Riemann zeta function and the $\sigma$-Riemann scatterer is achieved with a semi-infinite ($N \rightarrow \infty$) array of weak ($d_0 \rightarrow 0$) and point-particle ($\delta(x - x_0\log(n))$) scatterers, its measurable implementation requires breaking such ideal constraints. First, we investigate the effect of $N$ on the reflectance by multiple scattering, $R = |\psi_S(x<0)|^2/|\psi_0|^2$, which is an experimentally measurable quantity. Figure 2a shows the reflectance map $R(N,\lambda)$ under the weak scattering condition ($d_0 = 0.001$), which exhibits converging behaviours with increasing $N$. The result in Fig. 2b demonstrates the successful observation of the first 40 nontrivial Riemann-zero wavelengths $\lambda^{0,p}$ ($p = 1, 2, \ldots 40$) with the scatterer number $N = 1000$, where $\lambda^{0,p} = 4\pi x_0/t^{0,p}$ is obtained from the $p$th nontrivial zeros $t^{0,p}$.

The increase in $N$ can be considered the evolution of a material resulting from the addition of a series of particles, which leads to subsequent changes in scattering[35]. Such an evolving model reveals a unique nature of nontrivial Riemann zeros: non-oscillatory suppression of the reflectance. Figure 2c illustrates the evolutions of the reflectance $R$ for different wavelengths near the second nontrivial Riemann-zero wavelength $\lambda^{0,2}$. The result shows that the reflectance decreases monotonically with increasing $N$ exactly at the wavelength of a nontrivial Riemann zero, which is further confirmed with the evolution of $R$ at other nontrivial zeros (Fig. 2d for $\lambda^{0,p}$ with $p = 1, 2, \ldots 40$). On the other hand, the evolution of $R$ oscillates with $N$ at other wavelengths. In terms of the evolving scattering model[35], the monotonic decrease of $R$ corresponds to the optimal suppression of the density fluctuation at a given $k$ of $S(k)$. Similar to the design of stealthy hyperuniformity using the evolving model[35] that suppresses the density fluctuation in the certain range of $|k| < K$, the (1/2)-Riemann scatterer provides the optimal suppression of the density fluctuation at a set of $k$ defined by the nontrivial Riemann zeros.

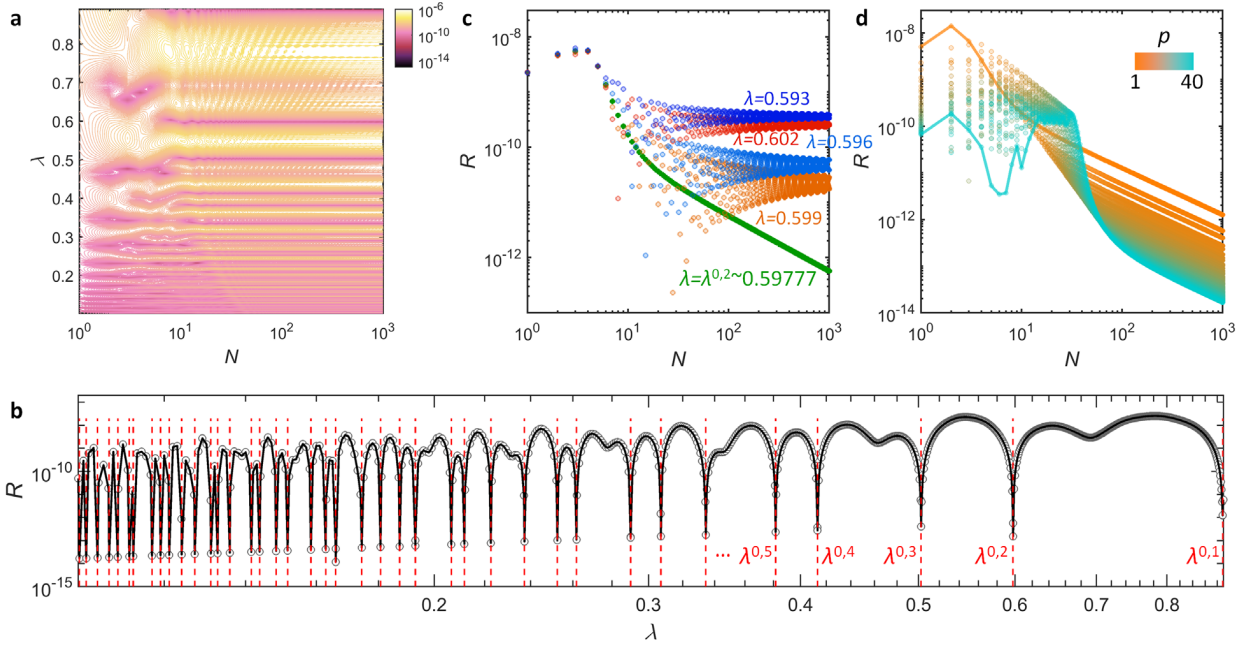

**Fig. 2. Evolution of the (1/2)-Riemann scatterer. a,** The multiple scattering reflectance $R$ as a function of the particle number $N$ and the wavelength $\lambda$. **b,** The multiple scattering reflectance $R$ (black solid line) as a function of $\lambda$ at $N = 1000$. Empty circles denote the reflectance from the first-order scattering. Red dashed lines indicate the wavelengths corresponding to nontrivial Riemann zeros $\lambda^{0,p}$ ($p = 1, 2, \ldots 40$). **c,** The evolutions of $R$ with respect to $N$ at the second nontrivial Riemann-zero wavelength $\lambda^{0,2} \sim 0.59777$ and its neighbouring wavelengths ($\lambda = 0.593$, 0.596, 0.599, and 0.602). **d,** The evolutions of $R$ with respect to $N$ at the nontrivial Riemann-zero wavelengths $\lambda^{0,p}$ ($p = 1, 2, \ldots 40$). The solid lines highlight the cases of $p = 1$ and 40. $x_0 = 1$, $d_0 = 0.001$, and $\psi_0 = 1$.

**Beyond the Born approximation**

Another important restriction in guaranteeing the validity of the measurement of Eq. (5) is the Born approximation. It is well known that the convergence of the Born series is valid when the spectral radius of the operator $M$ is less than unity, which can be achieved with weak scattering potentials[36]. This condition is determined by $d_0$, $N$, and the size of each scatterer[27]. While preserving the point-particle scatterer configuration, Fig. 3 illustrates the effect of $d_0$ and $N$ on the multiple scattering reflection for different Riemann-zero wavelengths. We note that the linear response between log($d_0$) and log($R$) guarantees the validity of the first-order Born approximation, which enables the light-based computing of nontrivial Riemann zeros. The increase of $d_0$ and $N$ breaks this approximation as widely examined in scattering theory[27,36]. Notably, the breaking occurs more rapidly at the lower-$p$ nontrivial Riemann zero with a larger $\lambda$. It is because an array of scatterers seems to be more clustered for the incidence of a larger $\lambda$, which breaks the constraint on the scatterer size despite the point shape of each scatterer.

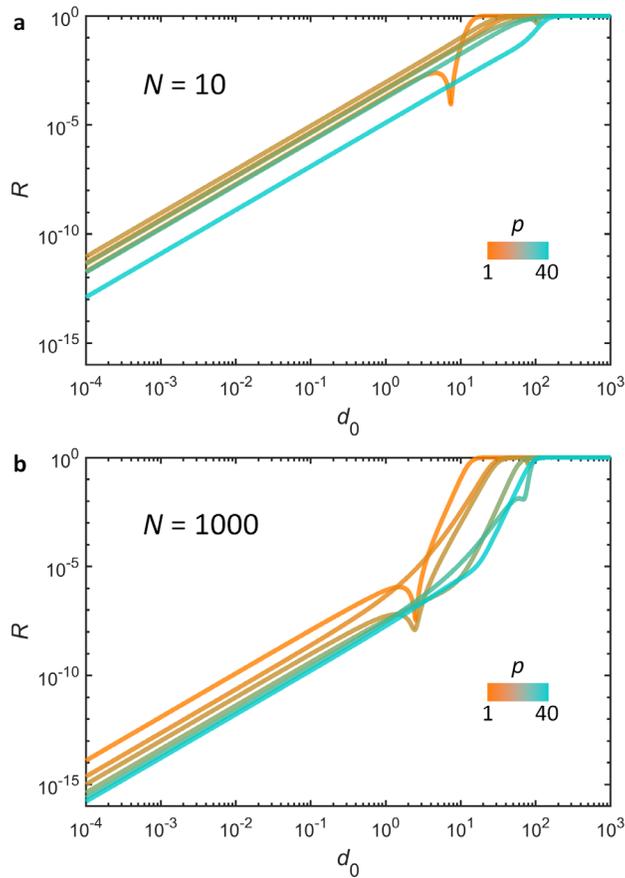

**Fig. 3. Beyond the Born approximation. a,b,** The multiple scattering reflectance $R$ as a function of $d_0$ at the wavelengths corresponding to nontrivial Riemann zeros $\lambda^{0,p}$ ($p = 1, 2, \ldots 40$): $N = 10$ (**a**) and $N = 1000$ (**b**). All the other parameters are the same as those in Fig. 2.

Extending the discussion on multiple scattering, we lift another critical restriction for practical implementation—the point particle condition. We replace each point particle of the $\sigma$-Riemann scatterer (Fig. 4a) with the corresponding layer of finite thickness (Fig. 4b), where the center of the layer is $x_n = x_0\log(n)$. The array of these layers is embedded in a homogeneous material of the potential $V_0 = k_0^2$. To reproduce the scattering from the ideal $\sigma$-Riemann scatterer, the thickness and potential of the $n$th layer are set to be $l_n = l_0 n^{-\sigma}$ and $V_n = [1 + (-1)^{n+1}\rho]^2 V_0$, respectively, where $l_0$ is the thickness of the first layer and $\rho$ is the coefficient of material inhomogeneity for two-phase multilayers (Fig. 4b). The designed layer corresponds to the effective realization of the point particle $d_n\delta(x - x_n)$, where $d_0 \sim (2k_0^2\rho)l_0$ for $d_n = (-1)^{n+1}d_0 n^{-\sigma}$ under weak scattering assumption (Supplementary Note S3).

Figure 4c shows the multiple-scattering reflectance from the multilayered Riemann scatterer with increasing $N$ at the wavelengths of nontrivial Riemann zeros ($p = 1, 5, 15$, and $30$). Notably, due to the breaking of the first-order Born approximation with $d_0 \sim (2k_0^2\rho)l_0$, the suppression of reflectance is weakened at shorter wavelengths, which correspond to higher-$p$ Riemann zeros. However, the relative suppression of reflectance at the Riemann-zero wavelengths compared to their nearby wavelengths significantly increases at higher $p$, as shown in the exemplified comparison among $p = 2, 15$, and $30$ in Supplementary Note S4. This observation is confirmed with the wavelength dependency of the discrepancy $\Delta R = R - |\psi_R|^2$ in Fig. 4d, which quantifies the contribution from higher-order scattering to the reflectance. As shown, the magnitude of $\Delta R$ maintains a relatively very low value at each Riemann-zero wavelength (red circles) when compared with the values of its nearby wavelengths. This result demonstrates that the ratio of higher-order scattering originating from the broken Born approximation is sufficiently small at the Riemann zeros, enabling the identification of the Riemann zeros even at higher $p$.

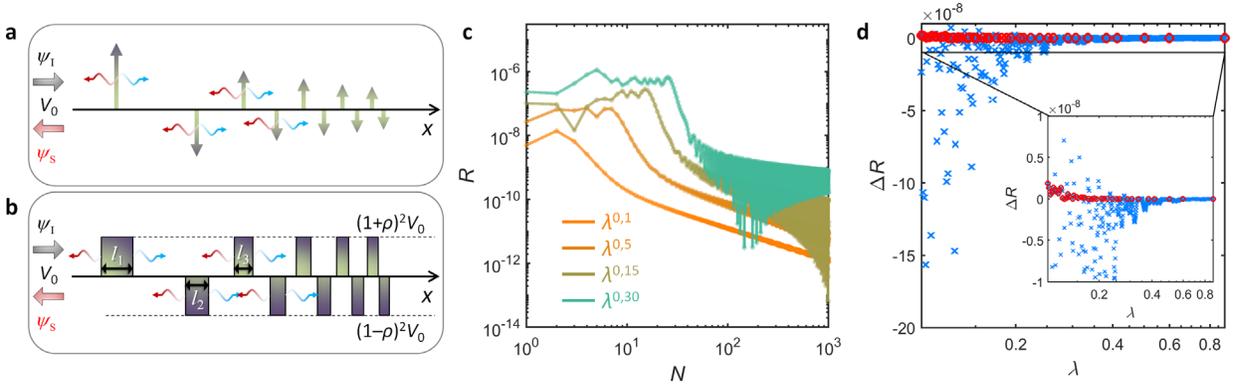

**Fig. 4. Multilayered Riemann scatterer. a,b,** Reproduction of the point-particle Riemann scatterer (**a**) with the multilayered one (**b**). **c,** The evolution of reflectance with respect to $N$ for the Riemann-zero wavelengths of $\lambda^{0,1}$, $\lambda^{0,5}$, $\lambda^{0,15}$, and $\lambda^{0,30}$. **d,** The difference between the multiple-scattering reflection and the first-order reflection, $\Delta R = R - |\psi_R|^2$, as a function of the wavelength $\lambda$. The red circles highlight the Riemann-zero wavelengths. The inset emphasizes a narrower range of $\Delta R$ for better visibility. The thickness coefficient and material inhomogeneity are set to be $l_0 = 0.01$ and $\rho = 10^{-3}$, respectively. All the other parameters are the same as those in Fig. 2.

## Discussion

The correspondence between the first-order scattering and the Riemann zeta function belongs to the class of wave-based mathematical operations, such as Fourier transforms using diffraction[37], matrix multiplications via interferometry[38], and integral equation solvers employing metamaterials[39]. Significantly, the relationship $\exp(2ik_0x_n) = n^{2ik_0x_0}$ for $x_n = x_0\log(n)$ demonstrates that logarithmically-spaced scatterers pave the way to the wave-based computing of other special functions, especially including variations of the Riemann zeta function involving complex variables. For example, the prime zeta function $P(s) = \sum_{p \in \text{primes}} \frac{1}{p^s}$ can be implemented in the same manner with the designed scatterers deposited only at the prime numbers $p$. The Hurwitz zeta function $\zeta(s,a) = \sum_{n=0}^{\infty} \frac{1}{(n+a)^s}$ requires the shifted positions of the particles $x_n = x_0\log(n+a)$ with the modified scattering amplitudes $d_n \sim (n+a)^{-\sigma}$. The reciprocal of the Riemann zeta function can also be realized with the designed scatterers in our approach, while the scattering amplitudes are modified according to the Möbius function $\mu(n)$. Such design freedom extends the application of wave-based computing into number theory.

According to our analysis of the broken Born approximation in the Riemann scatterer, maintaining dominant first-order scattering is critical. Although the straightforward approach involves weak perturbation and subwavelength design as we demonstrated, additional design flexibility still remains in the layer design using unnatural material parameters with metamaterials and some mathematical modifications such as using a suitable preconditioner[36]. Because the designed preconditioner provides the convergent form of the Born series even under large and strong perturbations, the stable calculation of multiple scattering with the modified Born series may reveal an intriguing connection to the infinite series of the Riemann zeta function.

In the realm of disordered photonics, the Riemann scatterer extends hyperuniformity with reciprocal-space shifts forms a class of logarithmically-spaced correlated disorder toward reflectionless design. Especially, the semi-infinite form of the Riemann scatterer introduces a unique mechanism for scattering suppression, distinct from antireflection coating, reflectionless potential profiles[40], reflectionless scattering modes[41], all of which primarily focus on finite-size scatterers. A comparable case would be parity-time-symmetric crystals for unidirectional transport[42], because nonuniform scattering amplitudes $d_n$ can be replaced with complex-valued wavenumbers $k_0$. The complete suppression of reflection in the (1/2)-Riemann scatterer with logarithmically-spaced scatterers also offers insights into the design of log-periodic antennas with directional radiation[43].

In conclusion, we developed the concept of the Riemann scatterer, which results in the scattering quantified by the Riemann zeta function under the first-order Born approximation. The Riemann hypothesis claims that completely zero reflection from the scatterer can be achieved only with a (1/2)-power-law distribution of scattering amplitudes. We examined the valid range of our theory by numerically investigating multiple scattering cases under the broken Born approximation. Further research is highly desirable to develop subsystem Hilbert-Pólya Hamiltonians for the Riemann scatterer, design other special functions, explore much higher zeros, and extend the theory to multiple scattering regimes, thereby bridging physics and number theory.

## Data availability

The data that support the plots and other findings of this study are available from the corresponding author upon request.

## Code availability

All code developed in this work will be made available upon request.

**Acknowledgements**

We acknowledge financial support from the National Research Foundation of Korea (NRF) through the Basic Research Laboratory (No. RS-2024-00397664), Young Researcher Program (No. 2021R1C1C1005031), and Midcareer Researcher Program (No. RS-2023-00274348), all funded by the Korean government. This work was supported by Creative-Pioneering Researchers Program and the BK21 FOUR program of the Education and Research Program for Future ICT Pioneers in 2024, through Seoul National University. We also acknowledge an administrative support from SOFT foundry institute.

**Author contributions**

All the authors conceived the idea, discussed the results, and contributed to the final manuscript.

**Competing interests**

The authors have no conflicts of interest to declare.

**Additional information**

**Correspondence and requests for materials** should be addressed to S.Y., X.P., or N.P.

**Figure Legends**

**Fig. 1. $\sigma$-Riemann scatterer. a,** A schematic of the $\sigma$-Riemann scatterer: a semi-infinite array of logarithmically spaced scatterers with a $\sigma$-power-law distribution of sign-alternating scattering amplitudes. The Riemann zeta function in the critical strip can be computed with the scattering wave $\psi_S$ for the incidence $\psi_I$. Red and blue curved arrows at each scatterer denote the scattering waves defined by the Green's function. **b,c,** Examples of scattering from the (1/2)-Riemann scatterer for $d_0 = 0.001$ (**b**) and $d_0 = 10$ (**c**). The arrows illustrate point scatterers. Red solid lines denote the multiple scattering $\psi_S$ calculated by the scattering matrix method[34] (Supplementary Note S2). Orange circles denote the first-order scattering $\psi_{S,1}$ obtained from Eq. (4). $N = 1000$, $x_0 = 1$ and $\psi_0 = 1$. **d,e,** The Riemann zeta function $\zeta(\sigma - 4\pi i x_0/\lambda)$ in the critical strip: amplitude (**d**) and phase (**e**) of $\zeta(\sigma - 4\pi i x_0/\lambda)$. The symbols '×' denote the first seven nontrivial Riemann zeros.

**Fig. 2. Evolution of the (1/2)-Riemann scatterer. a,** The multiple scattering reflectance $R$ as a function of the particle number $N$ and the wavelength $\lambda$. **b,** The multiple scattering reflectance $R$ (black solid line) as a function of $\lambda$ at $N = 1000$. Empty circles denote the reflectance from the first-order scattering. Red dashed lines indicate the wavelengths corresponding to nontrivial Riemann zeros $\lambda^{0,p}$ ($p = 1, 2, \ldots 40$). **c,** The evolutions of $R$ with respect to $N$ at the second nontrivial Riemann-zero wavelength $\lambda^{0,2} \sim 0.59777$ and its neighbouring wavelengths ($\lambda = 0.593, 0.596, 0.599$, and $0.602$). **d,** The evolutions of $R$ with respect to $N$ at the nontrivial Riemann-zero wavelengths $\lambda^{0,p}$ ($p = 1, 2, \ldots 40$). The solid lines highlight the cases of $p = 1$ and 40. $x_0 = 1$, $d_0 = 0.001$, and $\psi_0 = 1$.

**Fig. 3. Beyond the Born approximation. a,b,** The multiple scattering reflectance $R$ as a function of $d_0$ at the wavelengths corresponding to nontrivial Riemann zeros $\lambda^{0,p}$ ($p = 1, 2, \ldots 40$): $N = 10$ (**a**) and $N = 1000$ (**b**). All the other parameters are the same as those in Fig. 2.

**Fig. 4. Multilayered Riemann scatterer. a,b,** Reproduction of the point-particle Riemann scatterer (**a**) with the multilayered one (**b**). **c,** The evolution of reflectance with respect to $N$ for the Riemann-zero wavelengths of $\lambda^{0,1}$, $\lambda^{0,5}$, $\lambda^{0,15}$, and $\lambda^{0,30}$. **d,** The difference between the multiple-scattering reflection and the first-order reflection, $\Delta R = R - |\psi_R|^2$, as a function of the wavelength $\lambda$. The red circles highlight the Riemann-zero wavelengths. The inset emphasizes a narrower range

of $\Delta R$ for better visibility. The thickness coefficient and material inhomogeneity are set to be $l_0 = 0.01$ and $\rho = 10^{-3}$, respectively. All the other parameters are the same as those in Fig. 2.

# Supplementary Information for "Computing Riemann zeros with light scattering"

Sunkyu Yu[1†], Xianji Piao[2§], and Namkyoo Park[3*]

[1]Intelligent Wave Systems Laboratory, Department of Electrical and Computer Engineering, Seoul National University, Seoul 08826, Korea

[2]Wave Engineering Laboratory, School of Electrical and Computer Engineering, University of Seoul, Seoul 08826, Korea

[3]Photonic Systems Laboratory, Department of Electrical and Computer Engineering, Seoul National University, Seoul 08826, Korea

E-mail address for correspondence: [†]sunkyu.yu@snu.ac.kr, [§]piao@uos.ac.kr, [*]nkpark@snu.ac.kr

**Note S1.** Derivation of the first-order scattering

**Note S2.** Multiple scattering from the point-particle Riemann scatterer

**Note S3.** Multiple scattering from the multilayered Riemann scatterer

**Note S4.** Suppressed Riemann-zero reflection in multilayered Riemann scatterers

**Note S1. Derivation of the first-order scattering**

When applying Eq. (3) to Eq. (1) in the main text, the first-order scattering for the incidence $\psi_\mathrm{I} = \psi_0\exp(ik_0x)$ is obtained as follows:

$$\psi_{\mathrm{S},1}(x) = \frac{id_0\psi_0}{2k_0}\int_{-\infty}^{+\infty}\sum_{n=1}^{N}\frac{(-1)^{n+1}}{n^\sigma}\delta(x'-x_n)e^{ik_0x'}e^{ik_0|x-x'|}dx' \qquad (S1)$$
$$= \frac{id_0\psi_0}{2k_0}\sum_{n=1}^{N}\frac{(-1)^{n+1}}{n^\sigma}e^{ik_0x_n}e^{ik_0|x-x_n|},$$

where $x_n = x_0\log(n)$. In the region between the $p$th and $(p+1)$th scatterers, $x_p \leq x < x_{p+1}$, Eq. (S1) is expressed as:

$$\psi_{\mathrm{S},1}(x) = \frac{id_0\psi_0}{2k_0}\left[\left(\sum_{n=1}^{p}\frac{(-1)^{n+1}}{n^\sigma}\right)e^{+ik_0x} + \left(\sum_{n=p+1}^{N}\frac{(-1)^{n+1}}{n^\sigma}e^{2ik_0x_n}\right)e^{-ik_0x}\right]. \qquad (S2)$$

Notably, $\exp(2ik_0x_n) = \exp(2ik_0x_0\log(n)) = n^{2ik_0x_0}$, which leads to Eq. (4) in the main text.

**Note S2. Multiple scattering from the point-particle Riemann scatterer**

To analyse multiple scattering from the Riemann scatterer composed of point particles, we utilize the scattering matrix method[1]. For the $n$th point particle located at $x_n = x_0\log(n)$ with the scattering amplitude $d_n = (-1)^{n+1}d_0 n^{-\sigma}$, we define the waves around the particle, as $\psi_n^{L+}$, $\psi_n^{R+}$, $\psi_n^{L-}$, $\psi_n^{R-}$, where '+' and '−' denote the forward and backward propagating directions, respectively, while 'L' and 'R' represent the left- and right-side of the scatterer, respectively (Fig. S1a). When the particle is embedded in a homogeneous material governed by the Helmholtz equation $d^2\psi/dx^2 + V_0\psi = 0$, where $V_0 = k_0^2$, the waves are connected through the following interface transfer matrix:

$$\begin{bmatrix} \psi_n^{R+} \\ \psi_n^{R-} \end{bmatrix} = \begin{bmatrix} 1+i\dfrac{d_n}{2k_0} & i\dfrac{d_n}{2k_0} \\ -i\dfrac{d_n}{2k_0} & 1-i\dfrac{d_n}{2k_0} \end{bmatrix} \begin{bmatrix} \psi_n^{L+} \\ \psi_n^{L-} \end{bmatrix}, \quad (S3)$$

which characterizes the boundary condition between the waves around the $d_n$-weighted one-dimensional (1D) delta function potential. The relationship between the waves around neighbouring particles is characterized by the following propagation transfer matrix:

$$\begin{bmatrix} \psi_{n+1}^{L+} \\ \psi_{n+1}^{L-} \end{bmatrix} = \begin{bmatrix} e^{ik_0(x_{n+1}-x_n)} & 0 \\ 0 & e^{-ik_0(x_{n+1}-x_n)} \end{bmatrix} \begin{bmatrix} \psi_n^{R+} \\ \psi_n^{R-} \end{bmatrix}. \quad (S4)$$

The stable calculation of the total scattering matrix for a finite number of particles is achieved through iterative multiplications of the reformulated interface and propagation transfer matrices[1]. With the boundary condition defined by the incident waves $\psi_I$ and $\psi_I^R$ in Fig. S1b, the total scattering matrix leads to the scattering waves $\psi_S$ and $\psi_S^R$ in Fig. S1b for the finite-size Riemann scatterer. When assigning $\psi_I^R = 0$ and employing the semi-infinite condition $N\to\infty$, the resulting $\psi_S$ corresponds to the multiple scattering generalization of $\psi_{S,1}(x;x<x_1)$ in Eq. (5) in the main text.

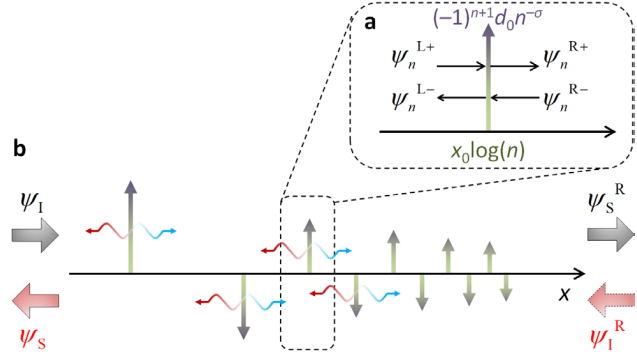

**Fig. S1. Multiple scattering from the point-particle Riemann scatterer. a,** Waves around the $n$th point particle, which is modelled by the weighted delta function potential. **b,** An array of point particles embedded in a homogeneous material. The incident waves are $\psi_I$ and $\psi_I^R$, and the scattering waves are $\psi_S$ and $\psi_S^R$. By setting $\psi_I^R = 0$, $\psi_S$ corresponds to multiple scattering reflection.

**Note S3. Multiple scattering from the multilayered Riemann scatterer**

To practically implement the suppressed reflection from the point-particle Riemann scatterer, we develop its approximated realization. Figure S2 illustrates the multilayer realization of the Riemann scatterer, which includes the unit layer design (Fig. S2a) and the multilayered structure (Fig. S2b). The $n$th point particle is approximately modelled with the $n$th layer that supports the potential $V_n = (k_0 + \delta k_n)^2$ and has a finite thickness $l_n$, which is embedded in the background material of the potential $V_0 = k_0^2$. Because $V_n$ is determined by the material parameter of the $n$th layer—for example, the permittivity of a simple medium—we focus on minimizing the number of values that $\delta k_n$ can take to decrease the number of necessary material phases.

While the center of the layer is $x_n = x_0 \log(n)$, the waves around the $n$th scattering layer in Fig. S2a are connected through the transfer matrix, as follows:

$$\begin{bmatrix} \psi_n^{R+} \\ \psi_n^{R-} \end{bmatrix} = \begin{bmatrix} e^{i(k_0+\delta k_n)l_n} + i\dfrac{\delta k_n^2 \sin((k_0+\delta k_n)l_n)}{2k_0(k_0+\delta k_n)} & i\dfrac{(2k_0+\delta k_n)\delta k_n}{2k_0(k_0+\delta k_n)} \sin((k_0+\delta k_n)l_n) \\ -i\dfrac{(2k_0+\delta k_n)\delta k_n}{2k_0(k_0+\delta k_n)} \sin((k_0+\delta k_n)l_n) & e^{-i(k_0+\delta k_n)l_n} - i\dfrac{\delta k_n^2 \sin((k_0+\delta k_n)l_n)}{2k_0(k_0+\delta k_n)} \end{bmatrix} \begin{bmatrix} \psi_n^{L+} \\ \psi_n^{L-} \end{bmatrix}. \quad (S5)$$

The relationship between the waves around neighbouring layers is determined by the following propagation transfer matrix:

$$\begin{bmatrix} \psi_{n+1}^{L+} \\ \psi_{n+1}^{L-} \end{bmatrix} = \begin{bmatrix} e^{ik_0[(x_{n+1}-x_n)-(l_{n+1}-l_n)]} & 0 \\ 0 & e^{-ik_0[(x_{n+1}-x_n)-(l_{n+1}-l_n)]} \end{bmatrix} \begin{bmatrix} \psi_n^{R+} \\ \psi_n^{R-} \end{bmatrix}. \quad (S6)$$

In calculating the multiple scattering from the multilayered Riemann scatterer, we employ Eqs. (S5) and (S6) directly to obtain rigorous solutions. However, to achieve scattering analogous to that from the point-particle Riemann scatterer, we need to design the parameters $l_n$ and $\delta k_n$ to derive the scattering analogous to that from a delta function potential. This design is achieved by comparing Eqs. (S3,S4) and (S5,S6) under suitable assumptions.

First, we neglect higher-order terms of $\delta k_n/k_0$ of second order or above by assuming the weak perturbation $\delta k_n/k_0 \ll 1$. The transfer-matrix equation in Eq. (S5), which will be compared with the interface transfer matrix in Eq. (S3), is then approximated as follows:

$$\begin{bmatrix} \psi_n^{R+} \\ \psi_n^{R-} \end{bmatrix} \sim \begin{bmatrix} e^{i(k_0+\delta k_n)l_n} & i\frac{\delta k_n}{k_0}\sin\left((k_0+\delta k_n)l_n\right) \\ -i\frac{\delta k_n}{k_0}\sin\left((k_0+\delta k_n)l_n\right) & e^{-i(k_0+\delta k_n)l_n} \end{bmatrix} \begin{bmatrix} \psi_n^{L+} \\ \psi_n^{L-} \end{bmatrix}. \quad (S7)$$

Equation (S7) can be transformed into the following scattering matrix formalism:

$$\begin{bmatrix} \psi_n^{R+} \\ \psi_n^{L-} \end{bmatrix} \sim e^{i(k_0+\delta k_n)l_n} \begin{bmatrix} 1-\left(\frac{\delta k_n}{k_0}\right)^2 \sin^2\left((k_0+\delta k_n)l_n\right) & i\frac{\delta k_n}{k_0}\sin\left((k_0+\delta k_n)l_n\right) \\ i\frac{\delta k_n}{k_0}\sin\left((k_0+\delta k_n)l_n\right) & 1 \end{bmatrix} \begin{bmatrix} \psi_n^{L+} \\ \psi_n^{R-} \end{bmatrix}$$

$$\sim e^{ik_0 l_n} \begin{bmatrix} 1 & i\frac{\delta k_n}{k_0}\sin(k_0 l_n) \\ i\frac{\delta k_n}{k_0}\sin(k_0 l_n) & 1 \end{bmatrix} \begin{bmatrix} \psi_n^{L+} \\ \psi_n^{R-} \end{bmatrix}, \quad (S8)$$

according to the condition $\delta k_n/k_0 \ll 1$. We also transform Eq. (S3) of a point-particle scatterer into the scattering matrix formalism, as follows:

$$\begin{bmatrix} \psi_n^{R+} \\ \psi_n^{L-} \end{bmatrix} = \frac{1}{1-i\frac{d_n}{2k_0}} \begin{bmatrix} 1 & i\frac{d_n}{2k_0} \\ i\frac{d_n}{2k_0} & 1 \end{bmatrix} \begin{bmatrix} \psi_n^{L+} \\ \psi_n^{R-} \end{bmatrix}. \quad (S9)$$

We assume the subwavelength condition for the layers, which allows for $\sin(k_0 l_n) \sim k_0 l_n$. The comparison between Eqs. (S8) and (S9) then leads to the relationship $l_n \sim d_n/(2k_0 \delta k_n)$. Because $d_n = (-1)^{n+1} d_0 n^{-\sigma}$, the multilayered Riemann scatterer requires the power-law distribution of the thickness $l_n = l_0 n^{-\sigma}$ with the relation $l_0 = (-1)^{n+1} d_0/(2k_0 \delta k_n)$. To obtain the positive-valued thickness $l_n > 0$, we utilize two-phase multilayers $\delta k_n = (-1)^{n+1} \rho k_0$, where $\rho$ is the constant coefficient for material inhomogeneity. The designed multilayer leads to the relationship between the effective weighting of the point particle $d_0$ and the initial thickness of the layer $l_0$: $d_0 = (2k_0^2\rho)l_0$.

Due to the relationship $d_0 = (2k_0^2\rho)l_0$, the effective weighting for a given $l_0$ depends on the wavelength of light $\lambda = 2\pi/k_0$. Because $d_0$ increases with a shorter wavelength $\lambda$, the Born approximation becomes more fragile for lower $\lambda$, which degrades the accuracy of light computing for higher nontrivial Riemann zeros, as shown in Fig. 4c,d in the main text.

The designed $l_n$ leads to the convergence of Eq. (S6) to the propagation transfer matrix in Eq. (S4). In the region between the $n$th and $(n+1)$th layers, the discrepancy between Eq. (S4) and Eq. (S6) is determined by:

$$\frac{l_{n+1} - l_n}{x_{n+1} - x_n} = \frac{d_0 \left[ (n+1)^{-\sigma} - n^{-\sigma} \right]}{2k_0^2 \rho x_0 \log\left(\frac{n+1}{n}\right)}. \tag{S10}$$

The Taylor expansion shows that Eq. (S10) asymptotically vanishes as $n$ increases, as follows:

$$\lim_{n \to \infty} \frac{l_{n+1} - l_n}{x_{n+1} - x_n} = \lim_{n \to \infty} \frac{-d_0 \sigma n^{-\sigma-1}}{2k_0^2 \rho x_0 \frac{1}{n}} = -\lim_{n \to \infty} \frac{d_0 \sigma}{2k_0^2 \rho x_0} \frac{1}{n^\sigma} = 0, \tag{S11}$$

for the values of $\sigma$ in the critical strip ($0 < \sigma < 1$). Therefore, the multilayered Riemann scatterer with $l_n = l_0 n^{-\sigma}$ and $\delta k_n = (-1)^{n+1}\rho k_0$ successfully provides the asymptotic matching between Eqs. (S3,S4) and (S5,S6).

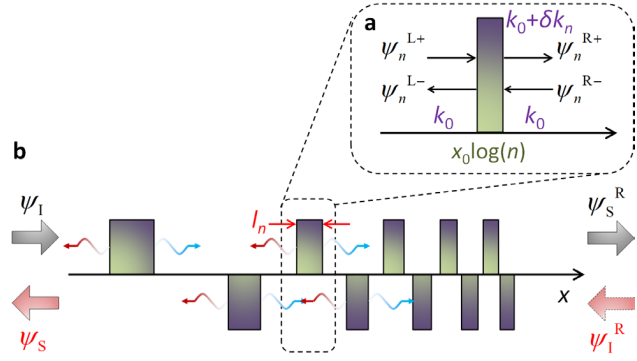

**Fig. S2. Multiple scattering from the multilayered Riemann scatterer. a,** Waves around the $n$th layer. **b,** An array of layers embedded in a homogeneous material. A scattering layer and a background material are characterized by their distinct optical wavenumbers, $k_0 + \delta k_n$ and $k_0$, respectively. The thickness of the $n$th layer is $l_n$. The incident waves are $\psi_I$ and $\psi_I^R$, and the scattering waves are $\psi_S$ and $\psi_S^R$. By setting $\psi_I^R = 0$, $\psi_S$ corresponds to multiple scattering reflection.

**Note S4. Suppressed Riemann-zero reflection in multilayered Riemann scatterers**

Figure S3 illustrates the evolutions of the reflectance $R$ for different wavelengths near the nontrivial Riemann-zero wavelengths $\lambda^{0,2}$ (Fig. S3a), $\lambda^{0,15}$ (Fig. S3b), and $\lambda^{0,30}$ (Fig. S3c). While Fig. S3a is very similar to the point-particle case (Fig. 2c in the main text), the reflectance $R$ increases at higher $p$ due to the breaking of the first-order Born approximation according to the relation $d_0 = (2k_0^2\rho)l_0$ (Fig. S3b,c). However, the discrepancy among the reflections at the Riemann-zero wavelength and its nearby wavelengths maintains large enough to identify the Riemann zeros, $\lambda^{0,15}$ and $\lambda^{0,30}$.

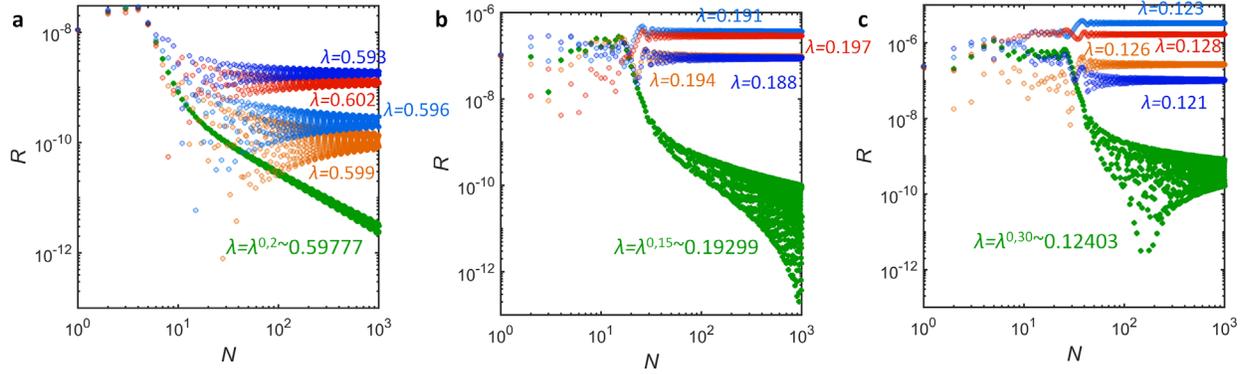

**Fig. S3. Riemann-zero reflections from multilayered Riemann scatterers. a-c,** The evolutions of $R$ with respect to $N$ at $\lambda^{0,2}$ (**a**), $\lambda^{0,15}$ (**b**), and $\lambda^{0,30}$ (**c**) and their neighbouring wavelengths. All the other parameters are the same as those in Fig. 4 in the main text.

**Supplementary References**